\documentclass[12pt]{article}
\usepackage{epsfig}
\usepackage{color}

\newcommand{\be}{\begin{equation}}
\newcommand{\ee}{\end{equation}}

\newcommand{\bea}{\begin{eqnarray}}
\newcommand{\eea}{\end{eqnarray}}
\newcommand{\beq}{\begin{equation}}
\newcommand{\eeq}{\end{equation}}
\newcommand{\nn}{\nonumber}

\def\la{\mathrel{\mathpalette\fun <}}

\def\fun#1#2{\lower3.6pt\vbox{\baselineskip0pt\lineskip.9pt
\ialign{$\mathsurround=0pt#1\hfil##\hfil$\crcr#2\crcr\sim\crcr}}}

\begin{document}

\title{
Central production of lepton-antilepton pairs
 and heavy quark composite states
in hadron diffractive collisions at ultrahigh energies}
\author{
V.V. Anisovich$^+$,
 M.A. Matveev$^+$,
V.A.  Nikonov$^{+ \diamondsuit}$,
J. Nyiri$^*$
}

\date{\today}

\maketitle

\begin{center}
{\it
$^+$National Research Centre "Kurchatov Institute",
Petersburg Nuclear Physics Institute, Gatchina, 188300, Russia}

{\it $^\diamondsuit$
Helmholtz-Institut f\"ur Strahlen- und Kernphysik,
Universit\"at Bonn, Germany}

{\it $^*$
Institute for Particle and Nuclear Physics, Wigner
RCP, Budapest 1121, Hungary }

\end{center}

\begin{abstract}

Central production of lepton-lepton pairs ($e^+e^-$ and $\mu^+\mu^-$)
and heavy quark composite states (charmonia and bottomonia)
in diffractive proton collisions (proton momenta transferred
${|\bf q_\perp}|\sim m/\ln s$)
are studied at ultrahigh energies ($\ln{s}>>1$), where
$\sigma_{tot}(pp^\pm)\sim\ln^N s$ with $1\la N\la 2$. The
$pp^\pm$-rescattering  corrections, which are not small, are
calculated in terms of the $K$-matrix approach modified for ultrahigh
energies. Two versions of hadron interactions are considered in detail:
the growth (i) $\sigma_{tot}(pp^\pm)\sim\ln^2 s$,
$\sigma_{inel}(pp^\pm)\sim\ln^2 s$ within the black disk mode
and (ii) $\sigma_{tot}(pp^\pm)\sim\ln^2 s$,
$\sigma_{inel}(pp^\pm)\sim\ln s$ within the resonant disk mode.
The energy behavior of the diffractive production processes differs
strongly for these modes, thus giving a possibility
to distinguish between the versions of the ultrahigh energy interactions.
\end{abstract}


\section{Introduction}

Recent measurements of the total, elastic and inelastic
$pp$ cross sections at the LHC \cite{Latino:2013ued,atlas,cms,alice},
and at cosmic ray energies by the Auger experiment \cite{auger},
reveal a successive step towards ultrahigh energy hadron physics.
For
$\sigma_{tot}(pp^\pm)$, $\sigma_{el}(pp^\pm)$ and
$\sigma_{inel}(pp^\pm)$ the data demonstrate a steady growth of the
type $\ln^N s$ with $1\la N\la 2$, similar to
that what was seen at preLHC
energies \cite{pre}, thus initiating a discussion about the
asymptotic regime, see
\cite{Halzen:2011xc,Uzhinsky:2011qu,Ryskin:2012az,Block:2012nj,ann1}
and references
therein. The data for the $pp$ diffractive scattering tell us that the
black spot appears in the impact parameter space, $\bf b$. It can be
an indication
of the beginning of the black disk regime but for a definite
confirmation a study of the diffractive cross sections at larger
energies, up to  $\sqrt{s}\sim 10^4$ TeV \cite{ann2}, is required.
The alternative can be the resonant mode, this regime starts also
with a black spot at small $b$ \cite{Anisovich:2014wha}.

In the search for and the recognition of asyptotics a study of diffractive production processes may be crucial. The
principal point for the study of production processes at ultrahigh energies is to take into account the rescattering corrections which
are large. The $K$-matrix technique modified for ultrahigh energies \cite{Anisovich:2014hca}
give us the possibility
to perform the corresponding calculations; we recall the main points
of the technique in Section 2.

We study the central productions of lepton-lepton pairs
($\ell\bar \ell=e^+e^-,\mu^+\mu^-,...$)
and heavy quark states $Q\bar Q= J/\psi, \Upsilon$, and so on:
\bea
\label{cp-1}
&&
pp\to p(\ell\bar \ell)p,
\\
&&
pp\to p(Q\bar Q)p.
\nn
\eea
In these two reactions the centrally produced particles,
leptons and heavy quarks, do not interact
strongly with incoming and outgoing  protons. But strong interactions
of protons are to be taken into account.
In Section 3,
in the framework of the black disk and resonant disk modes for ultrahigh energy hadron interactions, we calculate amplitudes of processes (\ref{cp-1}).

\begin{figure}
\centerline{
\epsfig{file=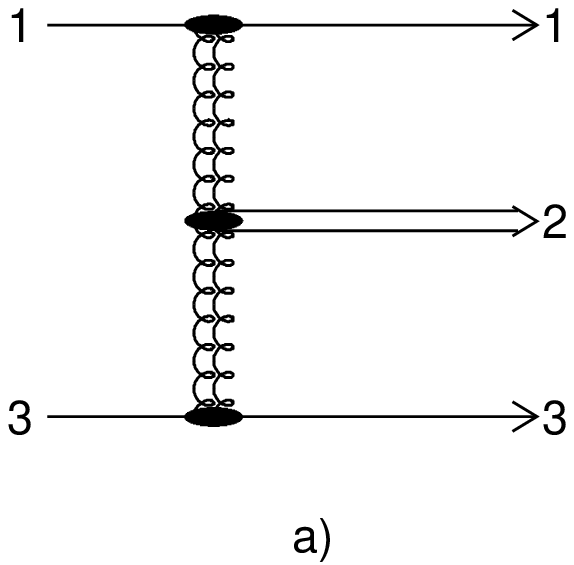,height=4.cm}\hspace{1.3cm}
\epsfig{file=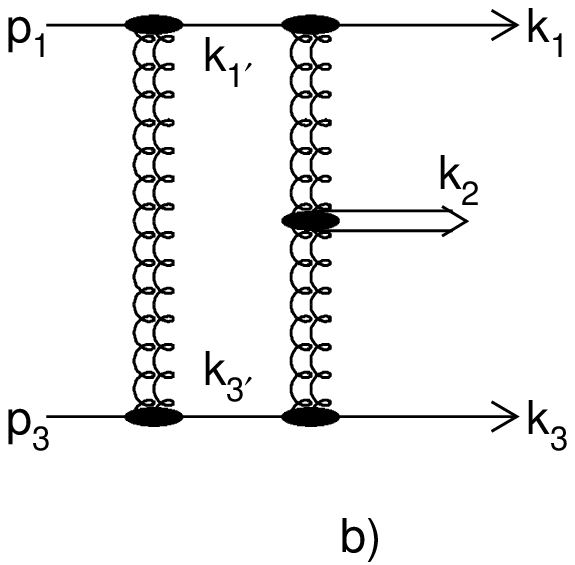,height=4.cm}}
\centerline{
\epsfig{file=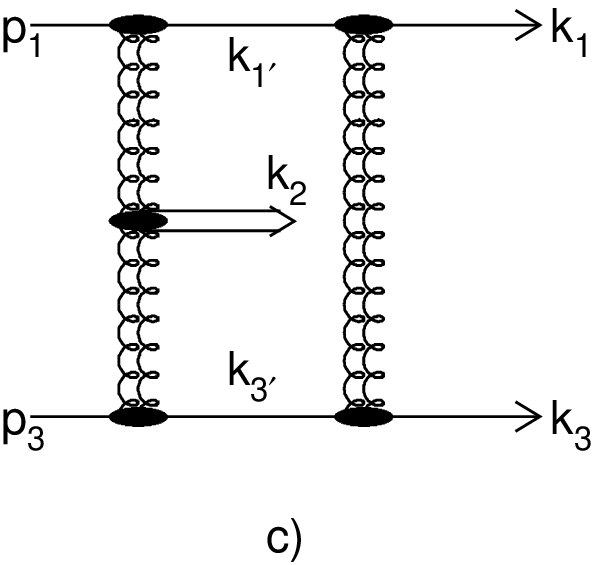,height=4.cm}\hspace{1.3cm}
\epsfig{file=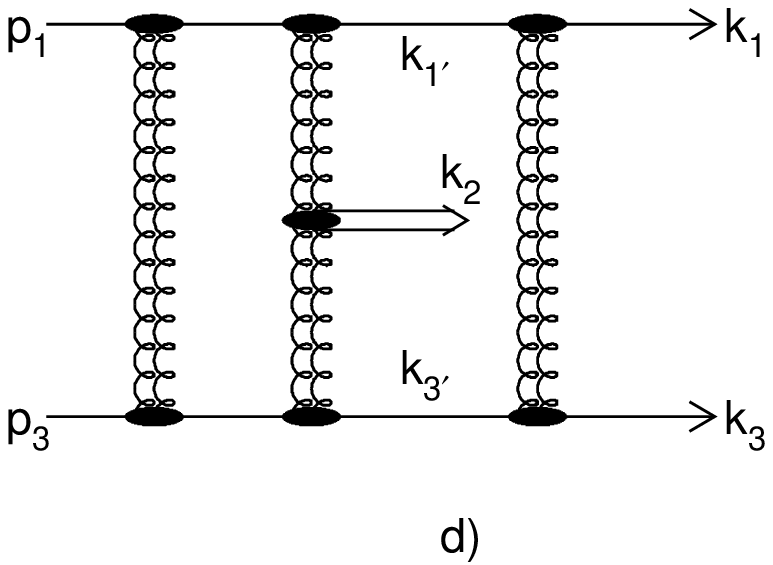,height=4.cm}}
\caption{a) Input diagram for diffractive production
$pp\to p(\ell\bar \ell)p$ or
$pp\to p(Q\bar Q)p$ , and b), c), d) diagrams
with subsequent rescatterings in initial and final states.
\label{fcp-1}}
\end{figure}

\section{Scattering amplitude and the $K(b)$-function}

Diffractive scattering amplitudes at ultrahigh energies
are usually considered in terms of the profile function $T(b)$
and the optical density $\chi(b)$. The $K$-matrix technique is
convenient for studying the production processes, see \cite{book3}
and references therein.
We use the following notation:
\bea \label{cp-2}
&&
\pi\frac{d\sigma_{el}}{d{\bf q}^2_\perp}=
a^2({\bf q}^2_\perp),\quad
a({\bf q}^2_\perp)=-\frac 12\int d^2{\bf b}\; e^{i{\bf b}{\bf q}_\perp} T(b)\,,
\\
&&
T(b)=1-\eta(b)\, e^{2i\delta(b)}=1-e^{-\frac12\chi(b)}
=
\frac{-2iK(b)}{1-iK(b)}=-2a({\bf b}^2,\ln s).
\nn
\eea
Here $b=|{\bf b}|$; in two-dimensional momenta transferred
we omit
the lower index $\perp$,
\i.e. ${\bf q}_\perp\to {\bf q}$; $a_{el}({\bf q}^2_\perp)$ is
the elastic scattering amplitude.
The profile function can be presented either in the standard form
using the inelasticity parameter $\eta(b)$ and the phase shift
$\delta(b)$, or in terms of the optical density $\chi(b)$ and the
$K$-matrix function $K(b)$. The $K$-matrix approach is based on the separation of the elastic rescatterings in the intermediate states:
the function $K(b)$ includes only the multiparticle states thus
being complex valued. The small value of the $ReA_{el}/ImA_{el}$
tells that $K(b)$ is dominantly imaginary.

\subsection{Eikonal approach for scattering amplitude
and the Feynman diagram technique}

For the scattering amplitude of hadrons
$A_{2\to 2}\Big((13)_{in}\to (13)_{out}\Big)$ the reproducing
integral reads:
\bea \label{cp-3}
&&
A_{2\to 2}\Big((13)_{in}\to (13)_{out}\Big)=
K_{2\to 2}\Big((13)_{in}\to (13)_{out}\Big)+
\\
&&
\int\frac{d^4k_{3'}}{(2\pi)^4i}\;
A_{2\to 2}\Big((13)_{in}\to 1'3'\Big)
\frac{1}{(m^2-k^2_{1'}-i0)(m^2-k^2_{3'}-i0)}\;
K_{2\to 2}\Big({1'}{3'}\to (13)_{out}\Big)\,. \nn
\eea
where $K_{2\to 2}$ is the block without two-particle states thus
being up to factor the $K$-matrix function; hadrons are denoted
by the indices $(1,3)$, the index $2$ we keep for the
centrally produced system, $(\ell\bar \ell)$ or $(Q\bar Q)$.

\subsubsection{Impact parameter presentation}

We consider the scattering amplitude in the cm-system
where
\be
p_1\equiv (p_0,{\bf p}_\perp,p_z)=(p+m^2/2p,0,p),
\quad
p_3=(p+m^2/2p,0,-p).
\ee
Therefore, we write:
\bea
\label{cp-4}
&&
{\bf k}_{1'\perp}+{\bf k}_{3'\perp}=0,\quad
{\bf k}_{1\perp}+{\bf k}_{3\perp}=0.
\\
&&
q^2_{3'}=(p_3-k_{3'})^2\simeq-{\bf k}^2_{3'\perp},\quad
q^2_{3'3}=(k_{3'}-k_{3})^2\simeq
-({\bf k}_{3\perp}-{\bf k}_{3'\perp})^2\,.
\nn
\eea
The $K$-matrix function $(-i)K(b)$ of the scattering amplitude
is real for the black disk regime. That means that the
mass-on-shell contributions are dominant in the loop diagrams.
For the rescattering diagrams this is realized by the
replacement
\bea \label{cp-5}
&&
\Big[(m^2-k^2_{1'}-i0)(m^2-k^2_{3'}-i0)\Big]^{-1}\to
-2\pi^2\delta(m^2-k^2_{1'})\delta(m^2-k^2_{3'})\,=\\
&&
-2\pi^2\delta\Big(k_{1'}^{(+)}k_{1'}^{(-)}
-(m^2+{\bf k}^2_{1'\perp})\Big)
\delta\Big(k_{3'}^{(+)}k_{3'}^{(-)}
-(m^2+{\bf k}^2_{3'\perp})\Big)\;, \nn
\eea
where
$k^{(+)}=k_0+k_z,\quad k^{(-)}=k_0-k_z$.
Then the right-hand side of Eq. (\ref{cp-3}) reads:
\bea \label{cp-6}
&&
A_{2\to 2}({\bf k}^2_{3\perp},\ln s)=
K_{2\to 2}\Big({\bf k}_{3\perp}^2,\ln s\Big)+
\\
&&
\int\frac{d^2k_{3'\perp}}{(2\pi)^2}
A_{2\to 2}({\bf k}^2_{3'\perp},\ln s) \frac{i}{4s}
K_{2\to 2}\Big(({\bf k}_{3'\perp}-{\bf k}_{3\perp})^2,\ln s\Big),
\nn
\eea
where $K_{2\to 2}/(4s) K$ is the $K$-matrix function in
momentum representation. Correspondingly, the Fourier transform
of it gives the $K$-matrix function in the
impact parameter space:
\bea
\label{cp-7}
&&
\frac{1}{4s}
K_{2\to 2}\Big({\bf k}^2_{\perp},\ln s\Big)=
\int d^2b\exp(i{\bf k}{\bf b})K({\bf b}^2,\ln s)\,,
\\
&&
\frac{i}{4s}
A_{2\to 2}\Big({\bf k}^2_{\perp},\ln s\Big)=
\int d^2b\exp(i{\bf k}{\bf b})a({\bf b}^2,\ln s)\,,
\nn
\eea
Equation (\ref{cp-6}) in the impact parameter space is written as:
\be \label{cp-8}
a({\bf b}^2,\ln s)=
iK({\bf b}^2,\ln s)+a({\bf b}^2,\ln s)\;
iK({\bf b}^2,\ln s)\,.
\ee
Thus, we have the formula of the eikonal approach:
\be \label{cp-9}
a({\bf b}^2,\ln s)=
\frac{iK({\bf b}^2,\ln s)}{1-iK({\bf b}^2,\ln s)},
\ee
see Eq. (\ref{cp-2}). The function $K({\bf b}^2,\ln s)$ depends on
the energy and realizes effectively the instantaneous interaction
which manifests itself in shrinking of diffractive cones with the energy increase.

\subsubsection{Analytical properties of the
diffractive scattering amplitude}

The scattering amplitude has $t$-channel singularities at $t=(\sum m)^2$; for the $pp$ scattering they are
$t=\mu_\pi^2,\ 4\mu_\pi^2,\ 9\mu_\pi^2$ and so on. All these singularities
are effectively far from the region of consideration of the amplitude:
at ultrahigh energies the amplitude depends on $\tau\sim|t|\ln^n s$ and
the singularities in the $\tau$-plane tend to infinity with the
energy increase, $\tau_{sing}=(\sum m)^2\ln^n s\to\infty$. As a
result, integrations over $k^{(+)},\,k^{(-)}$ are factorized thus
transforming the amplitude (\ref{cp-3}) into a set of loop diagrams:
\be \label{cp-10}
a({\bf b}^2,\ln s)=
\sum\limits_{\ell=1}^\infty\bigg(iK({\bf b}^2,\ln s)\bigg)^\ell ,
\ee
that reproduces (\ref{cp-9}).

The analytical properties in the $s$-plane are determined by the
loop diagram and corresponding cut discontinuities - that are
logarithmic terms, $A_{2\to 2}\sim s\ln^N s $. At ultrahigh energies
one should take into account the $u$-channel cut as well. For the
positive signature we write
$s\ln^N s +u\ln^N u=s\ln^N s +(-s)\ln^N(-s) \sim i\pi\ln^{N-1} $
that gives the dominant imaginary part. The choice of the $K(b)$
in accordance with Eq. (\ref{cp-2}) takes into
account this property.

\begin{figure}[ht]
\centerline{\epsfig{file=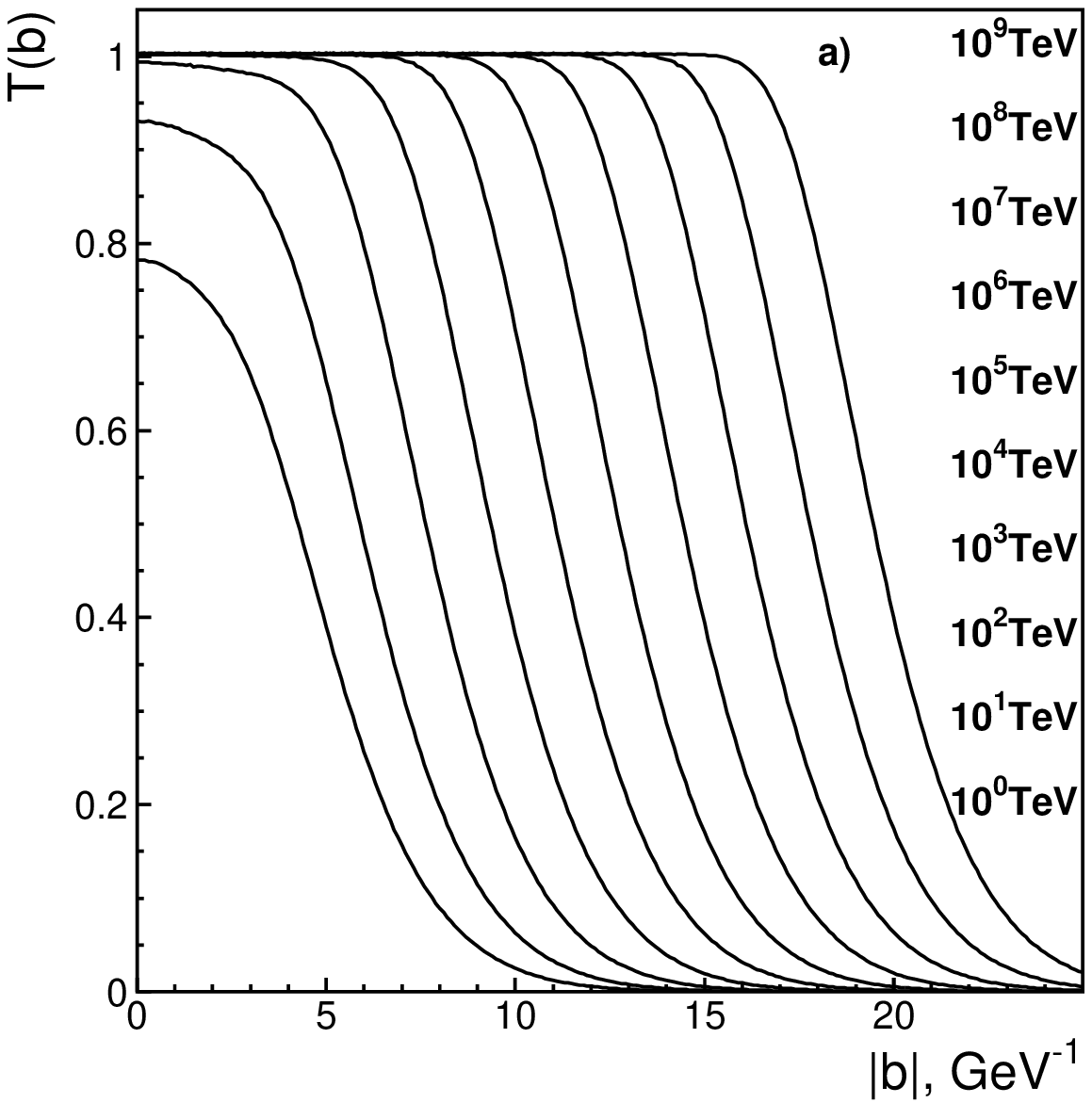,width=7cm}
           \epsfig{file=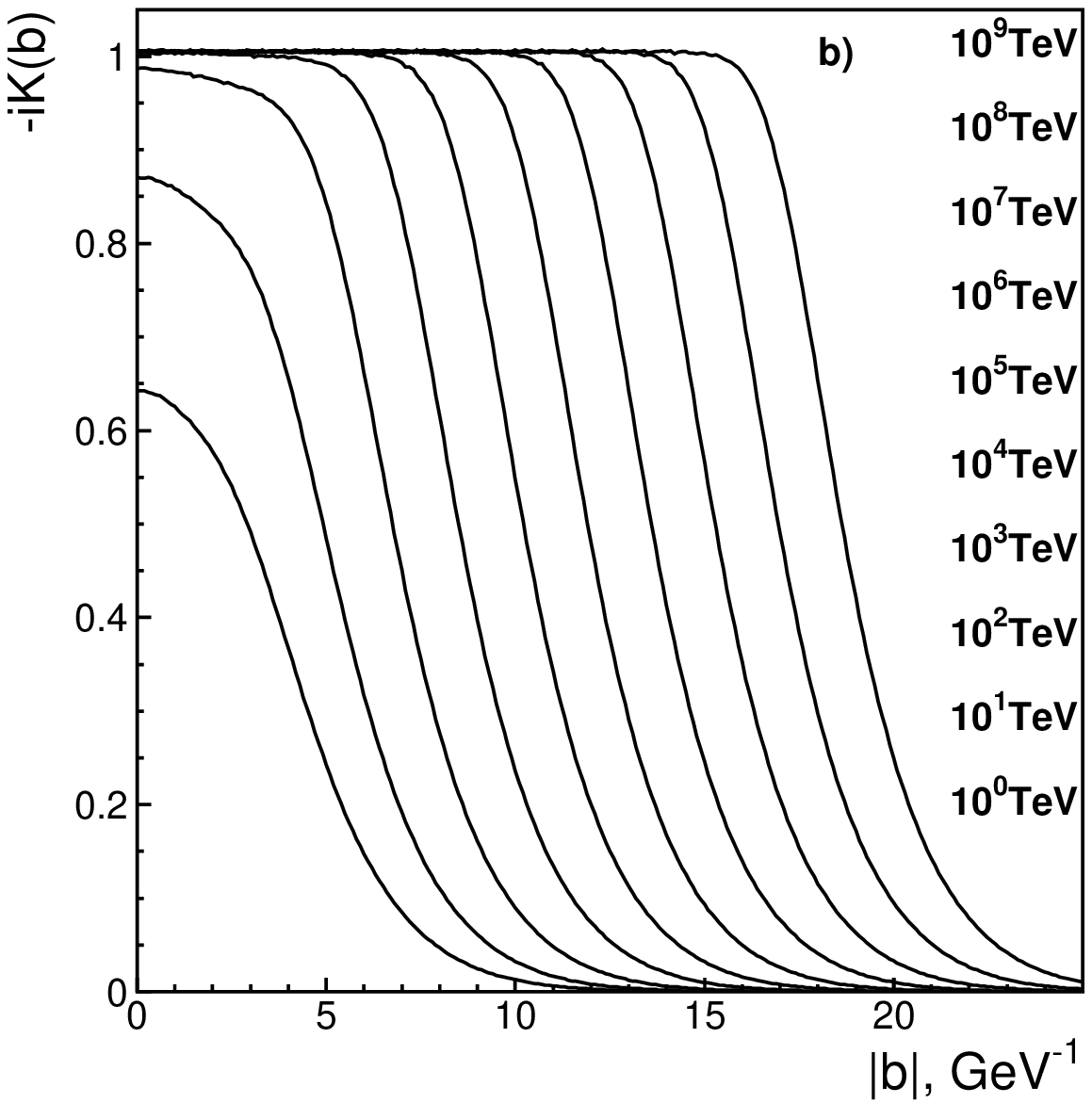,width=7cm}            }
\caption {Black disk mode:
 a)  Profile functions, $T(b)$,
 at  $\sqrt{s}=1,\,10,\,10^2,... 10^{9}$ TeV
with  $T(b)\to 1$ at $b<R \ln s$
 and b) corresponding $K$-matrix function determined as
 $T(b)= -2iK(b)/[1-iK(b)]$.
\label{fcp-2}}
\end{figure}

\subsection{Black disk and resonant disk modes}

We know that the profile function $T(b)$
reaches the black disk limit at small impact parameters,
$b\la 0.5\;fm$ at LHC energies. But it is not known
whether $T(b)$ is frozen at $T(b)=1$ or continues to
increase with the energy growth \cite{Anisovich:2014wha}. Because of that
we consider two
versions for the asymptotic behavior: (i) with the black disk
regime, $T(b)\to 1$ at $b<R_{disk}$, and (ii) with the maximal
value of the profile function corresponding to the resonant
disk regime, $T(b)\to 2$ at $b<R_{disk}$.

\subsubsection{Black disk limit in terms of the Dakhno-Nikonov model}

The Dakhno-Nikonov model \cite{DN} demonstrates us a
representative example of application of the optical density
technique for the consideration of $pp^\pm$ collisions
at ultrahigh energies when $\ln s>>1$. In the model
the black disk is formed by the pomeron cloud and rescatterings
are described within the eikonal approach. The same model may
demonstrate the reformulation to the $K$-matrix technique.

The behavior of amplitudes at ultrahigh energies is
determined by leading complex-j singularities, in the
Dakhno-Nikonov model that are leading and next-to-leading pomerons
with trajectories
$\alpha({\bf q}^2)\simeq 1+\Delta-\alpha'{\bf q}^2$.
The fit of refs. \cite{ann1,ann2} gives
$\Delta= 0.27 $ and $\alpha'_P= 0.12$ GeV$^{-2}$.

In terms of the $K$-matrix approach the black disk mode means the
assumed freezing of the $-iK(b)$ in the interaction area:
 \bea \label{23-17}
\Big[-i K(b)\Big]_{\xi\to\infty}\to 1
 \qquad && {\rm at} \; b<R_0 \, \xi\,,
 \\
  \Big[-iK(b)\Big]_{\xi\to\infty}\to 0
\qquad &&
 {\rm at} \; b>R_0 \, \xi \, ,\nn
 \\
  \xi=\ln\frac{s}{s_R},\quad s_R\simeq 6.4\; 10^3\;{\rm GeV}^2,
 && {\rm with}\;
 R_{0}\simeq 2\sqrt{\alpha'\Delta}\simeq 0.08 \; {\rm fm}.\nn
\eea
The growth of the radius of the black disk is slow: the
small value of $R_0$ is caused by the large mass of glueballs
\cite{AIP-conf,ijmp} and the effective mass of gluons
\cite{parisi,field}.
The black disk mode results in
\bea
&&
\sigma_{tot}\simeq 2\pi(R_0\xi)^2,
\\
&&
\sigma_{el}\simeq \pi(R_0\xi)^2,\quad
\sigma_{inel}\simeq \pi(R_0\xi)^2.
\nn
\eea
For the black disk radius the corrections of the order of $\ln \xi$
exist $R_{black\; disk}\simeq R_0\xi+\varrho\ln\xi$ but
they become apparent in the Dakhno-Nikonov model at energies
of the order of the Planck mass, $\sqrt{s}\sim 10^{17}$ TeV.

\begin{figure}
\centerline{
            \epsfig{file=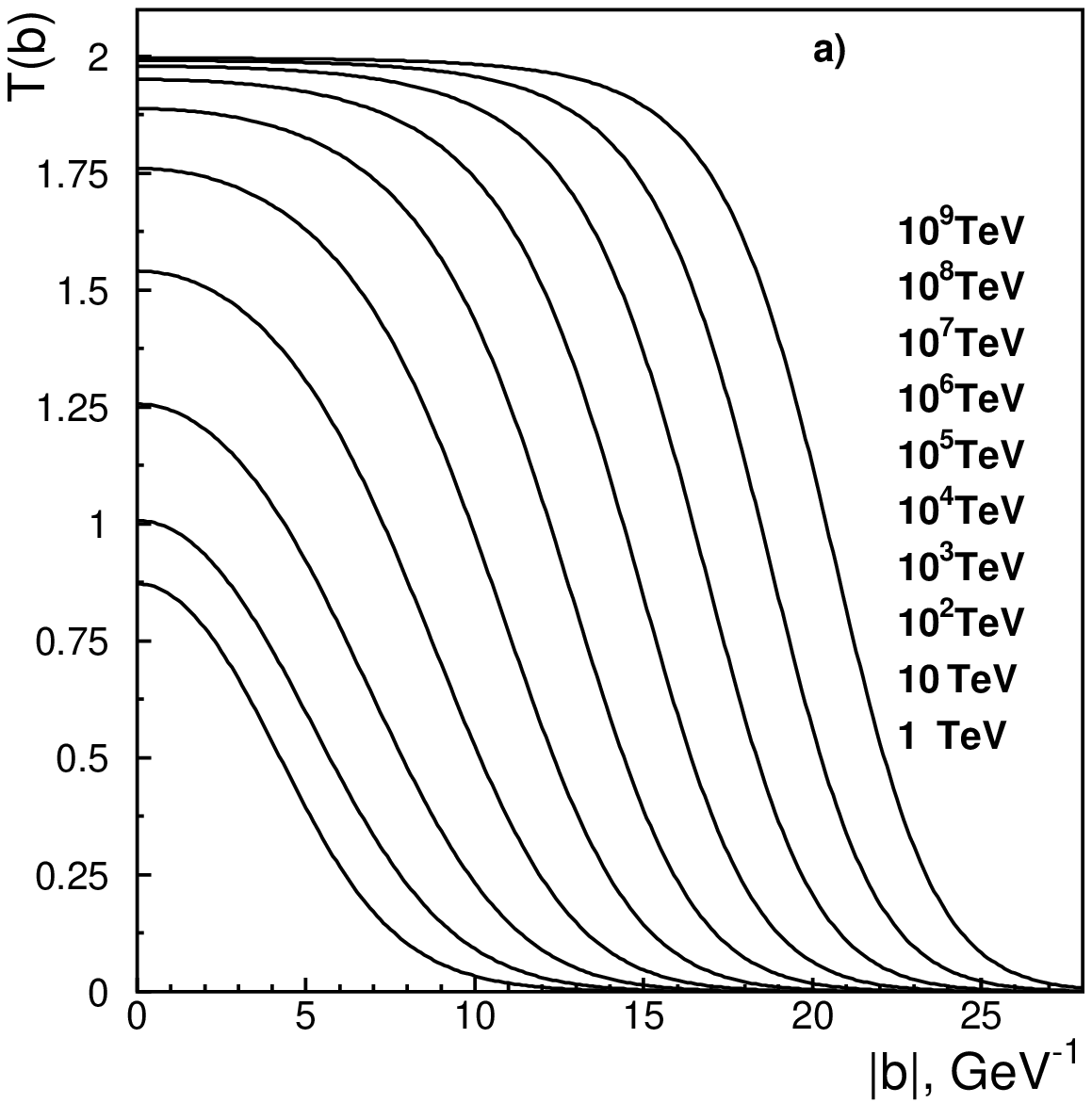,width=7cm}
            \epsfig{file=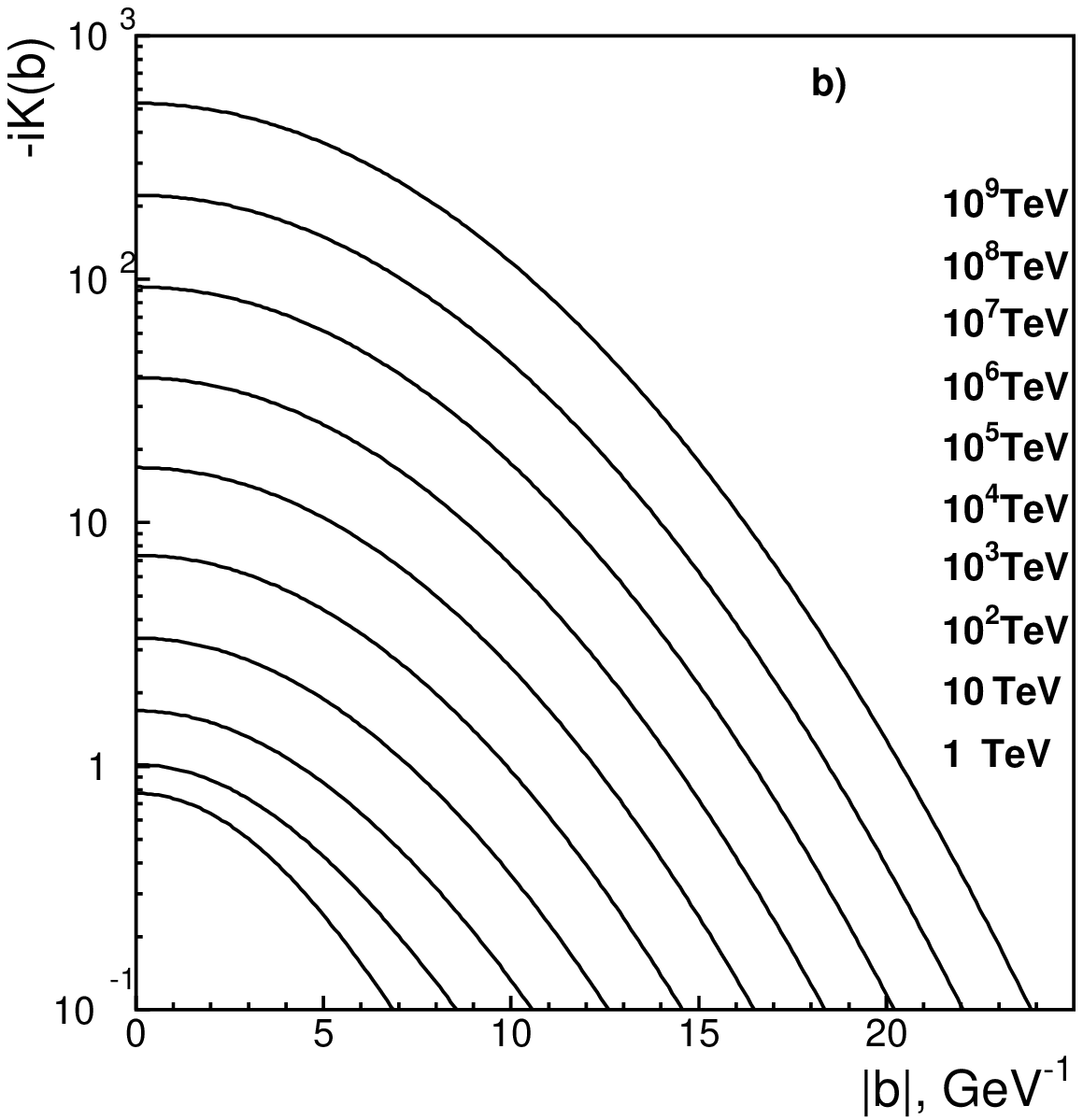,width=7cm}}
\caption{
Resonant disk mode:
a) the profile function $T(b)$
and b) $K$-matrix function, $-iK(b)$,
$[-iK(b)]_{\xi\to \infty}\to \infty$ at
$b< R_0\xi$.
\label{fcp-3}}
\end{figure}

\subsubsection{Resonant disk and the $K$-matrix function growth }

From the data it follows that both $T(b)$ and $-i K(b)$
are increasing with energy, being less
than unity. If the eikonal mechanism does not quench the
growth, both characteristics cross the black disk limit getting
 $T(b)>1$, $-i K(b)>1$. If $-i K(b)\to \infty$ at $\ln s\to \infty$,
which corresponds to a growth caused by the supercritical
pomeron ($\Delta>0$), the diffractive scattering process gets to
the resonant disk mode.

For following the resonant disk switch-on we use the two-pomeron
model with parameters providing the description of data at 1.8 TeV and
7 TeV, namely:
\bea
\label{rd5}
&&
-i K(b)=\int \frac{d^2q}{(2\pi)^2}
\exp{\Big(-i{\bf q}{\bf b}\Big)}
\sum g^2 s^\Delta
e^{-(a+\alpha\xi){\bf q}^2)}
\\
&&
= \sum
\frac{g^2}{4\pi(a+\alpha'\xi)}
\exp{\Big[\Delta\xi-\frac{{\bf b}^2}{4(a+\alpha'\xi)}\Big]}\,,
\qquad \xi=\ln \frac {s}{s_0}.
\nn
\eea
The following parameters are found
for the leading and the next-to-leading pomerons:
\be
\begin{tabular}{l|l|l}
  parameters       & leading pole  & next-to-leading    \\
\hline
$\Delta$                    &  0.20          &  0      \\
$\alpha'_P$ [GeV$^{-2}$] &  0.18           &  0.14   \\
$a$   [GeV$^{-2}$]     &  6.67           & 2.22   \\
$g^2$  [ mb ]        &  1.74           &  28.6  \\
$s_0$  [GeV$^{2}$]   &  1               &   1   \\
\end{tabular}
\ee
The resonant interaction regime occurs at
$b<2\sqrt{\alpha'\Delta}\xi=R_{rd}\xi$, in this region $T(b)\to 2$.
In terms of the inelasticity parameter and the phase shift
it corresponds to $\eta\to 1$ and $\delta\to \pi/2$.
Cross sections at $\xi\to\infty$ obey
$\sigma_{tot}\simeq 4\pi R^2_{rd} \xi^2$, $\sigma_{el}/\sigma_{tot}\to 1$
and $\sigma_{inel}\simeq 2\pi R_{rd} \xi$.

At the energy $\sqrt s\sim 10$ TeV the cloud constituents fill out the
proper hadron domain, the region $\leq 1$ fm, and that happens in
both modes.
It is demonstrated in Figs. \ref{fcp-2}a, \ref{fcp-3}a where it is seen
that the profile functions $T(b)$ coincide practically in both modes as
well as the K-functions $-iK(b)$.
Differences appeared at $\sqrt s\sim 1000$ TeV: $T(b)\simeq 1.5$
at $b\la 0.5$ fm and the black zone has shifted to
$b\simeq 1.0-1.5$ fm, Figs. \ref{fcp-2}b, \ref{fcp-3}b. With further
energy increase the radius of the black band increases as
$2\sqrt{\Delta\alpha'}\xi\equiv R_{rd}\xi$. The rate of growth
in both modes is determined by the leading singularity and the fit
of the data in the region $\sqrt s\sim 1-10 $ TeV gives approximately
the same values of $\Delta$ and $\alpha'$ for both cases thus
providing $R_{rd}\simeq R_0$.

\section{Production amplitude: screening effects due to
initial and final state rescatterings}

Return now to the productions of centrally produced particles,
$\ell\bar\ell$ and $Q\bar Q$. The problem we solve is to calculate
effects of the rescatterings
presuming that the input amplitude is known. Therefore, we calculate
an amplitude prolongation into the region of ultrahigh energies supposing we know the amplitude at lower energies. The evolution of
the amplitude we calculate is determined by the growth of the hadron disk size, its long-range component.

 \subsubsection{Input amplitude for production of
 $\ell\bar\ell$ and $Q\bar Q$}

The input amplitude for the production of three particles
is shown in Fig. \ref{fcp-1}a, it is written as:
\be
\phi_{0}({\bf k}^2_{1},\xi_{12};\;{\bf k}^2_{3},\xi_{23})
=\int d^2b_{1}d^2b_{3}f_{0}(b_{1},\xi_{12}\;;b_{3}\,,\xi_{23})
\exp\Big(i{\bf k}_{1}{\bf b}_{1}+i{\bf k}_{3}{\bf b}_{3}
\Big).
\label{08-9}
\ee

For the black disk mode we write the input term as:
\bea
\label{23_28}
{\rm {\bf k}-space:}&&
\phi_{0}({\bf k}^2_{1},\xi_{12};\;{\bf k}^2_{3},\xi_{23})
=g_{2\to 3}
a({\bf k}^2_1,\xi_{12})\,a({\bf k}^2_3,\xi_{23})\,,
\\
{\rm {\bf b}-space:}&&\quad
f_{0}(b_{1},\xi_{12}\;;b_{3}\,,\xi_{23})
=\frac14g_{2\to 3}
 T(b_1,\xi_{12})\, T(b_3,\xi_{23})\,
\nn
\eea
with $a({\bf k}^2,\xi)$ and $T(b,\xi)$ being determined by
 Eq. (\ref{cp-2}).

In the resonant disk mode the diffractive processes are determined
by pomeron-type exchanges, therefore we use the two-pomeron term.
We write in the momentum and impact parameter spaces,
correspondingly:
\bea
\label{23_26}
{\rm  {\bf k}-space:
}&&\quad
\phi_{0}({\bf k}^2_{1},\xi_{12};{\bf k}^2_{3},\xi_{23})
=g_{2\to 3}
\exp{\Big[\Delta\xi_{12}-\alpha'\xi_{12}{\bf k}^2_1\Big]}
\exp{\Big[\Delta\xi_{23}-\alpha'\xi_{23}{\bf k}^2_3\Big]}\,,
\\
{\rm {\bf b}-space:
}&&
\quad
f_{0}(b_{1},\xi_{12}\;;b_{3},\xi_{23})
=g_{2\to 3}
\frac{e^{
\Delta\xi_{12}}}{4\pi\alpha'\xi_{12}}
\exp\Big[-\frac{{\bf b}^2_1}{4\alpha'\xi_{12}}\Big]
\frac{e^{
\Delta\xi_{23}}}{4\pi\alpha'\xi_{23}}
\exp\Big[-\frac{{\bf b}^2_3}{4\alpha'\xi_{23}}\Big].
\nn\eea

\subsubsection{Initial state rescatterings}

Rescatterings in the initial state give additional terms
into the production amplitude. The one-rescattering term reads:
\bea
\phi_{1}({\bf k}^2_{1},\xi_{12};\;{\bf k}^2_{3},\xi_{23})
&=&\int d^2b_{1}d^2b_{3}
\; iK(b,\xi)f_{0}(b_{1},\xi_{12}\;;b_{3}\,,\xi_{23})
\exp\Big(i{\bf k}_{1}{\bf b}_{1}+i{\bf k}_{3}{\bf b}_{3}
\Big),
\nn
\\
\xi&=&\xi_{12}+\xi_{23},\quad {\bf b}={\bf b}_{1}+{\bf b}_{3}.
\label{08-10}
\eea
In the impact parameter space the rescattering results in factor
$iK(b,\xi)$. The two-rescatterings term contains the factor
$(iK(b,\xi))^2$ and so on.
The summation of all terms
$\sum\limits_{n=0,1,2,...}f_n$ generates a standard $K$-matrix factor
$(1-iK(\xi,b))^{-1}$
and we write for the input term corrected by taking into account
the initial state interactions:
\be  \label{cp-19}
\sum
\limits_{n=0}^\infty\phi_n({\bf k}^2_{1},\xi_{12};\;{\bf k}^2_{3},\xi_{23})
=\int d^2b_{1}d^2b_{3}
\; \frac{1}{1-iK(\xi,b)}
f_0(b_{1},\xi_{12}\;;b_{3}\,,\xi_{23})
\exp\Big(i{\bf k}_{1}{\bf b}_{1}+i{\bf k}_{3}{\bf b}_{3}
\Big).
\ee
The final state interactions lead to the same factor, and we have
finally:
\be  \label{cp-20}
f(b_{1},\xi_{12}\;;b_{3}\,,\xi_{23})=
 \frac{f_0(b_{1},\xi_{12}\;;b_{3}\,,\xi_{23})}{\Big(1-iK(b,\xi)\Big)^2}
 .
\ee
Factor $\Big(1-iK(b,\xi)\Big)^{-1}$ is universal for taking into account
the rescattering corrections.

Rescattering corrections behave differently at ultrahigh energies:
for the black disk mode
$[1-iK(b)]^{-1}\to 1/2$ at $\sqrt{s}\to\infty$
while for the resonant mode
$[1-iK(b)]^{-1}\to 0$ at $\sqrt{s}\to\infty$.

\section{Generating operator for production amplitude}

One can write Eq. (\ref{cp-20}) by
the operator
\bea
&&
f_0(b',\xi'\;;b''\,,\xi'')
\frac{\partial}{\partial\Big(iK(b,\xi)   \Big)},\\
&&
\xi'+\xi''=\xi,\quad {\bf b'}+{\bf b''}={\bf b}
\nn
\eea
acting on the scattering amplitude $a({\bf b}^2,\xi)$ given in
Eqs. (\ref{cp-9}), (\ref{cp-10}). For
the reactions investigated here the generating operator
looks as a plaything but it can be really helpful when the
productions of the $\ell\bar \ell$ or
$Q\bar Q$ pairs are considered in multihadron reactions like that
studied in \cite{Anisovich:2014hca}.

\vspace{3mm}
\centerline{***}

In terms of the modified $K$-matrix technique we consider central
production $pp\to p(\ell\bar\ell) p$ or
$pp\to p(Q\bar Q)p$ when momenta transferred to protons are small,
${\bf k}^2_\perp \sim m^2/\ln^2 s$. Rescattering corrections, which
are calculated in straightforward way, lead to substantial differences
in energy behavior of production amplitudes at different modes,
the black disk and resonant disk ones. In the resonant disk mode
the scattering correction factor $[1-iK(b)]^{-1}$ decreases
with energy growth $[1-iK(b)]^{-1}\to 0$ at $\sqrt{s}\to\infty$
while for the black disk mode
$[1-iK(b)]^{-1}\to 1/2$ at $\sqrt{s}\to\infty$; energy behavior
differentials emphasize importance of studies of production processes.

\subsubsection*{Acknowledgment}

We thank  M.G. Ryskin
and A.V. Sarantsev for useful discussions and comments.
 The work was supported by grants RFBR-13-02-00425 and
 RSGSS-4801.2012.2.

\end{document}